\documentclass[letterpaper, 10 pt, conference]{ieeeconf}  % Comment this line out if you need a4paper

\IEEEoverridecommandlockouts                              % This command is only needed if 
\usepackage{graphics} % for pdf, bitmapped graphics files
\usepackage{epsfig} % for postscript graphics files
\usepackage{mathptmx} % assumes new font selection scheme installed
\usepackage{times} % assumes new font selection scheme installed
\usepackage{amsmath} % assumes amsmath package installed
\usepackage{amssymb}  % assumes amsmath package installed
\usepackage{amsfonts}
\usepackage{graphicx}
\usepackage{multirow}
\usepackage{tikz}
\usepackage{cite}
\usepackage{romannum}
\usepackage{mathrsfs}
\usepackage{caption}
\usepackage[ruled,linesnumbered,noend]{algorithm2e}
\usepackage{hyperref}

\DeclareMathAlphabet{\mathcal}{OMS}{cmsy}{m}{n}

\usepackage{amsthm}

\newtheorem{remark}{Remark}
\title{\LARGE \bf
Towards Improving the Performance of the RNN-based Inversion Model in Output Tracking Control
}

\author{Shengwen Xie$^{1}$ and Juan Ren$^{1,\dagger}$% <-this % stops a space
\thanks{$^{1}$S. Xie and $^{1}$J. Ren are with the Department of Mechanical Engineering,
        Iowa State University, Ames, IA 50011, USA 
        {\tt\small swxie@iastate.edu},{\tt\small juanren@iastate.edu}}%
\thanks{$^{\dagger}$ Corresponding author.}
}

\begin{document}

\maketitle
\thispagestyle{empty}
\pagestyle{empty}
\setlength{\textfloatsep}{-0.1cm}
\setlength{\floatsep}{-0.1cm}

%%%%%%%%%%%%%%%%%%%%%%%%%%%%%%%%%%%%%%%%%%%%%%%%%%%%%%%%%%%%%%%%%%%%%%%%%%%%%%%%
\begin{abstract}
With the advantages of high modeling accuracy and large bandwidth, recurrent neural network (RNN) based inversion model control has been proposed for output tracking. However, some issues still need to be addressed when using the RNN-based inversion model. First, with limited number of parameters in RNN, it cannot model the low-frequency dynamics accurately, thus an extra linear model has been used, which can become an interference for tracking control at high frequencies. Moreover, the control speed and the RNN modeling accuracy cannot be improved simultaneously as the control sampling speed is restricted by the length of the RNN training set. Therefore, this article focuses on addressing these limitations of RNN-based inversion model control. Specifically, a novel modeling method is proposed to incorporate the linear model in a way that it does not affect the existing high-frequency control performance achieved by RNN. Additionally, an interpolation method is proposed to double the sampling frequency (compared to the RNN training sampling frequency). Analysis on the stability issues which may arise when the proposed new model is used for predictive control is presented along with the instructions on determining the parameters for ensuring the closed-loop stability. Finally, the proposed approach is demonstrated on a commercial piezo actuator, and the experiment results show that the tracking performances can be significantly improved.
\end{abstract}

%%%%%%%%%%%%%%%%%%%%%%%%%%%%%%%%%%%%%%%%%%%%%%%%%%%%%%%%%%%%%%%%%%%%%%%%%%%%%%%%
\section{Introduction}
Piezo actuators (PEAs) have been widely used for nanopositioning related applications such as investigating the mechanic properties of materials based on atomic force microscope (AFM)\cite{mollaeian2018nonlinear}, aeroelastic control of aircraft wings \cite{wang2019active}, and vibration control \cite{gedikli2019active}. Even though there have been many models developed to capture the dynamics of PEAs, the challenge of controlling piezo actuators still lies in modeling the nonlinearities---hysteresis and creep effects over large frequency bandwidth and amplitude span\cite{gu2016modeling}. 

Recently, neural networks have been proposed to model the dynamics of PEAs \cite{cheng2015neural,liu2016inversion}. However, the feedforward neural network is not able to the utilize the sequence information of the input \cite{cheng2015neural,liu2016inversion}. In contrast, the input to the recurrent neural network (RNN) is a time series, thus it is more suitable for modeling the dynamical system and can achieve any level of modeling accuracy given enough number of parameters \cite{schafer2007recurrent}. In our previous work, RNN has been proposed for modeling PEA systems \cite{xie2019recurrent}. Since the obtained RNN is equivalent to a nonlinear state space model, a nonlinear predictive controller has been proposed to realize output tracking \cite{xie2019recurrent}. However, the computation load is very heavy which limits the prediction horizon and further improvement of the control precision.

Alternatively, RNN can be used to model the inversion dynamics of the PEA system \cite{xie2019tracking}. Compared to some existing models, such as the ferromagnetic material hysteresis \cite{cao2013inversion}, Prandtl-Ishlinskii hysteresis inversion model \cite{shan2012dual}, RNN-based inversion model (RNNinv) can achieve large modeling bandwidth and high accuracy. Moreover, since the inversion model is used without an extra controller, it avoids the heavy computation burden for real-time applications. Nevertheless, it shows that RNNinv cannot fully account for the low-frequency dynamics and time-varying dynamics of PEAs due to the length limit of the RNN training set, thus a linear model embedded with an error term (LME) has been proposed to make a predictive controller for trajectory tracking at low frequencies \cite{xie2019tracking}. Although LME is expected to account for the low-frequency dynamics only, it will interfere with the control contribution of the RNNinv at high frequencies resulting in an increase of the tracking errors\cite{xie2019tracking}. Also, due to the limited length of the training set, the sampling frequency of RNN can not be set very high. Therefore, this article is motivated to overcome these two issues of RNNinv for PEA output tracking.

 The main contribution of this work is the development of a novel modeling approach such that the dynamics of the system is separated based on the frequency ranges. Specifically, for a linear system model $M$, its dynamics is sperated into two linear models $A$ and $B$ in such a way that the high-frequency dynamics of $M$ is characterized by $A$ while the low-frequency dynamics by $B$. In terms of this work, let $M$ denote the dynamics of RNNinv+PEA, it is expected that LME accounts for the low-frequency dynamics ($A$) of the RNNinv+PEA system while the high-frequency dynamics ($B$) is close to 1 (i.e., $M(j\omega)=1$ for large $\omega$) given that RNNinv can efficiently account for the PEA high frequency dynamics. Another contribution is that for the RNN trained with the sampling frequency of $f_s$, an interpolation method is proposed to make it able to be operated at the sampling frequency of $2f_s$. The reason for doing this is that for some applications such as high-speed AFM imaging, the sampling frequency can affect the spatial resolution of the obtained images thus high sampling frequency is preferred \cite{ren2014high,xie2019high}.

To realize precision trajectory tracking, a predictive controller based on the model $M$ is used. Note that $M$ has parameters which differentiate between ``high-frequency'' and ``low-frequency''. It turns out that these parameters can affect the stability of the predictive controller. In other words, apart from the parameters of the predictive controller, the parameters of $M$ should also be chosen carefully to ensure the closed-loop stability. It should be noted that it is quite difficult to analyze the closed-loop stability of MPC. Although sometimes prediction horizon $N_h$ cannot be tuned to stabilize the system \cite{muller2017quadratic}, generally speaking, large $N_h$ can ensure the closed-loop stability especially for linear systems. Many theoretical results account for a general system and thus quite conservative \cite{limon2006stability}, or focus on setpoint stabilization \cite{reble2012unconstrained} thus not applicable for general reference tracking problem. In this work, we analyze the closed-loop stability of the unconstrained predictive control problem. Results about the closed-loop stability as well as how they are used to determine the relevant parameters are also presented.

%The rest of the paper is organized as follows. Preliminaries about the application of RNN-based inversion model in output tracking are given in Section \Romannum{2}. The new modeling method and the interpolation approach of doubling the sampling frequency are presented in Section \Romannum{3}. In Section \Romannum{4}, the closed-loop stability of the predictive controller is investigated. Experiment and simulation results are given in Section \Romannum{5}. Finally, Section \Romannum{6} concludes the paper.

%On the other hand, these results on reference tracking often assume that there exists a reference dynamics i.e., the reference signal can be represented by $r_{k+1}=f(r_k)$, in case the reference cannot described by the above dynamics, these results cannot be used\cite{kohler2018nonlinear}.

\section{Preliminaries}
\begin{figure}[b!]
\centering
\captionsetup{justification=centering}
\vspace{0.4cm}
\includegraphics[scale=0.7]{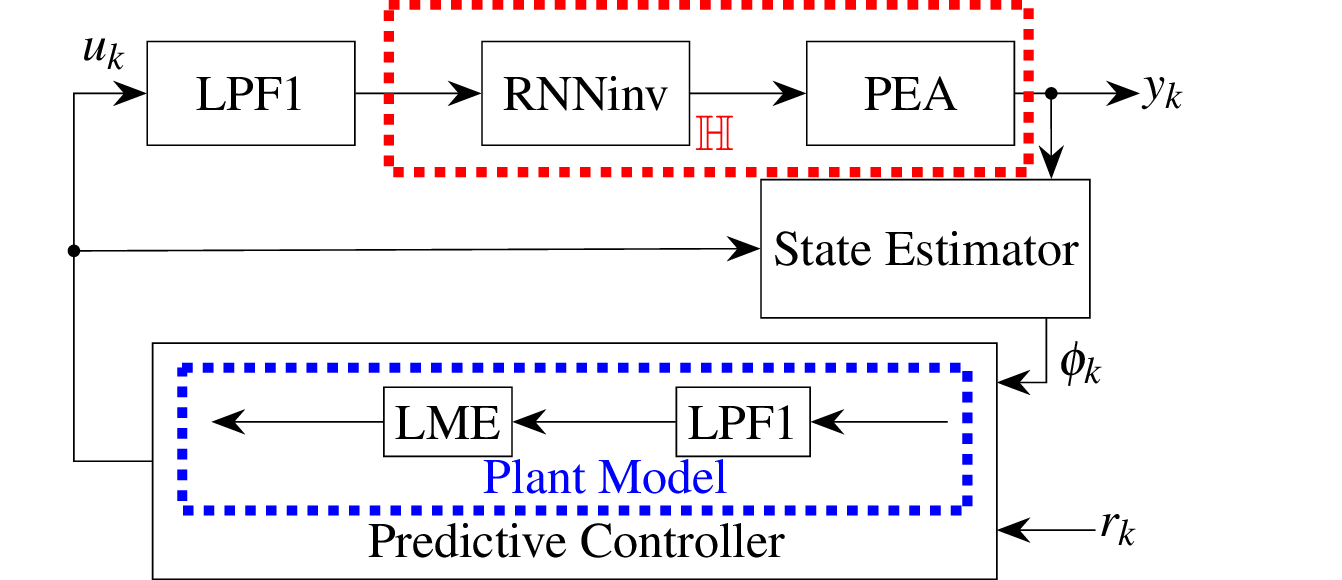}
\caption{Schematic block diagram of using RNNinv for PEA output tracking.}
\label{figRNN}
\end{figure}

The RNN-based inversion model (RNNinv) can be represented by the follow state space model,
\begin{equation}
\begin{aligned}
x_{k+1}&=\tanh(W_1x_k+B_2+B_1u_{(r),k})\\
y_{(r),k}&=W_2x_k+B_3
\end{aligned},
\label{EqRNN}
\end{equation}
where $x_k\in \mathbb{R}^{N\times 1}$, $u_{(r),k}$, and $y_{(r),k}$ are the system states, input and output at the sampling instant $k$, respectively. The diagram of the controller for output tracking is shown in Fig. \ref{figRNN}. As can be seen in Fig. \ref{figRNN}, RNNinv is cascaded with the PEA system forming the system $\mathbb{H}$ \cite{xie2019tracking}. Since RNNinv can account for the high-frequency dynamics accurately with the low-frequency dynamics partly modeled, $\mathbb{H}$ is modeled by the following linear model embedded with an error term (LME) to account for the time-varying and low-frequency dynamics,

\begin{equation}
\begin{aligned}
\zeta_{k+1}&=A_e\zeta_k+B_e\hat{u}_k+G_e\hat {e}_k\\
\hat {y}_k&=C_e\zeta_k
\end{aligned}\text{,}
\label{LME}
\end{equation}
where $\hat{u}_k$ is the input to the LME, $\hat{e}_k=y_k-\hat{y}_k$ is the model output error with $y_k$ as the actual PEA output. The problem is that the high-frequency dynamics of LME will interfere with that of $\mathbb{H}$ resulting in modeling control errors. With the new modeling method in the next section, the high-frequency dynamics of LME will not affect that of $\mathbb{H}$.

%thus, the frequency response of $\mathbb{H}$  in the high-frequency region should be 1. Although LME is designed to deal with the low-frequency dynamics, its high-frequency dynamics will interfere with that of $\mathbb{H}$  resulting in a modeling error for the high-frequency region. Next, we show how to model $\mathbb{H}$ such that its high-frequency dynamics is 1 while the low-frequency dynamics are the same as that of LME.

\section{Model the System Dynamics}
\subsection{Incorporating the LME}
Suppose we have a system $G$ such that its dynamics in the frequency range $[f_h,f_s/2]$ coincides with that of system $G_1$ and the dynamics in $[0,f_l]$ can be characterized by that of system $G_2$, the goal is to represent $G(j\omega)$ with $G_1(j\omega)$ and $G_2(j\omega)$. Note that $f_l \geq f_h $. Instead of solving the problem directly, we introduce extra dynamics $G_e$ and show that $G_eG$ can be represented by $G_1$, $G_2$ and $G_e$ with the following proposition. Here, we use $G_eG$ to denote the dynamics of the system combining $G_e$ and $G$.
$G$, $G_1$, and $G_2$ can be expressed in state-space forms with Eq. (\ref{eqG}), Eq. (\ref{eqG1}), and Eq. (\ref{eqG2}), respectively.

\begin{equation}
\begin{aligned}
x_{g,k+1}&=f(x_{g,k},u_{g,k})\\
y_{g,k}&=h(x_{g,k})
\end{aligned}
\label{eqG}
\end{equation}

\begin{equation}
\begin{aligned}
x_{g1,k+1}&=f_1(x_{g1,k},u_{g1,k})\\
y_{g1,k}&=h_1(x_{g1,k})
\end{aligned}
\label{eqG1}
\end{equation}

\begin{equation}
\begin{aligned}
x_{g2,k+1}&=f_2(x_{g2,k},u_{g2,k})\\
y_{g2,k}&=h_2(x_{g2,k})
\end{aligned}
\label{eqG2}
\end{equation}

\textbf{Proposition 1.} Assume $G$ satisfies 
\begin{equation}
f(a+b,c+d)=\bar{f}(a,c)+\hat{f}(b,d)
\label{CondSCD}
\end{equation}

\begin{figure}[t!]
\centering
\captionsetup{justification=centering}

\includegraphics[trim={9.5cm 19.5cm 10.5cm 4cm},scale=0.7]{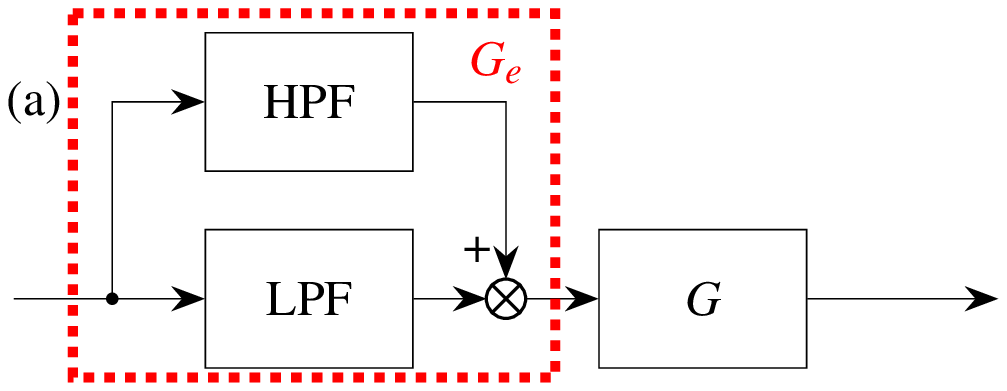}

\includegraphics[trim={9.5cm 19.5cm 10.5cm 4cm},scale=0.7]{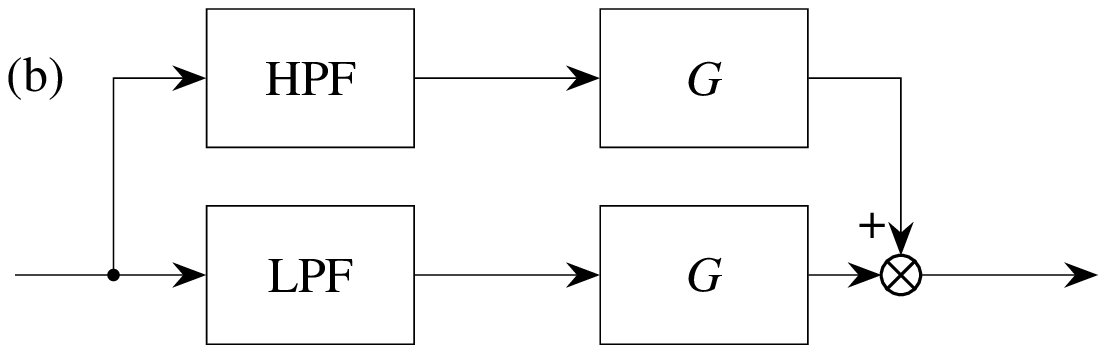}

\includegraphics[trim={9.5cm 19.5cm 10.5cm 4cm},scale=0.7]{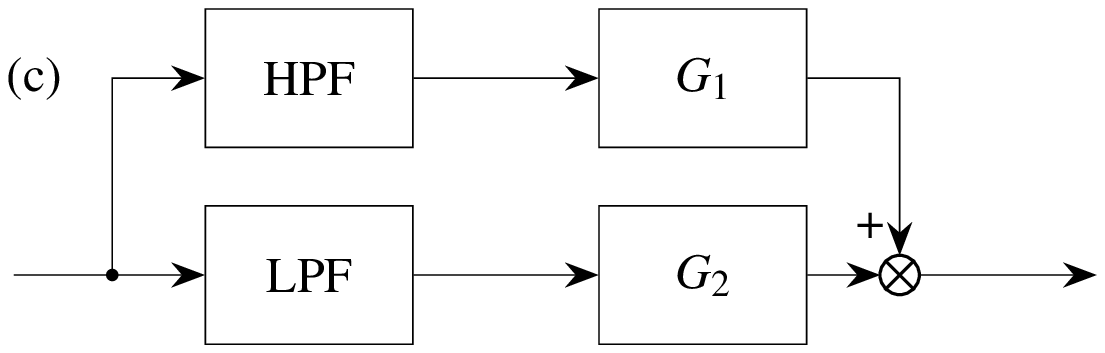}

\caption{Combine two models.}
\label{combine}
\end{figure}

$G_e$ is constructed with a parallel connection of a high-pass filter (HPF) (Eq. (\ref{HPF})) and a low-pass filter (LPF) (Eq. (\ref{LPF})) as shown in Fig. \ref{combine}(a). 

\begin{equation}
\begin{aligned}
\alpha_{k+1}&=\bar{A}\alpha_{k}+\bar{B}\bar{{u}}_k\\
\bar{z}_k&=\bar{C}\alpha_k
\end{aligned}
\label{HPF}
\end{equation}

\begin{equation}
\begin{aligned}
\beta_{k+1}&=\underline{A}\beta_{k}+\underline{B}\underline{{u}}_k\\
\underline{z}_k&=\underline{C}\beta_k
\end{aligned}
\label{LPF}
\end{equation}

$G_1$ and $G_2$ are described as above. Then $G_eG$ can be represented by
\begin{equation}
\begin{aligned}
\begin{bmatrix}
\alpha_{k+1}\\
x_{g1,k+1}\\
\beta_{k+1}\\
x_{g2,k+1}
\end{bmatrix}
=&\begin{bmatrix}
\bar{A}\alpha_{k+1}+\bar{B}\bar{{u}}_k\\
f_1(x_{g1,k},~\bar{C}\alpha_k)\\
\beta_{k+1}+\underline{B}\underline{{u}}_k\\
f_2(x_{g2,k},~\underline{C}\beta_k)
\end{bmatrix}\\
{z}_{g,k}=&h_1(x_{g1,k})+h_2(x_{g2,k})
\end{aligned}\text{.}
\label{GeG}
\end{equation}

\renewenvironment{proof}{\textbf{Proof.}}{$\square$}
\begin{proof}
$G_e$ is shown in Fig. \ref{combine}(a). With the assumption Eq.~(\ref{CondSCD}), Fig. \ref{combine}(a) is equivalent to Fig. \ref{combine}(b). The HPF can be tuned to only allow signals in the frequency range $[f_h,f_s/2]$ to pass and similar for LPF: only signals in the frequency range $[0,f_l]$ can pass. Then Fig. \ref{combine}(b) is equivalent to Fig. \ref{combine}(c) which can be expressed using Eq. (\ref{GeG}). This completes the proof. 
\end{proof}

\begin{remark}
The cutoff frequencies $f_{cL}$ (LPF) and $f_{cH}$ (HPF) should be chosen such as that the signals at all the frequencies can pass through $G_e$. If the system is linear, Eq. (\ref{CondSCD}) would be satisfied automatically. If Eq. (\ref{CondSCD}) does not hold, there will be modeling errors for the proposed modeling approach.
\end{remark}

\begin{figure}[t!]
\centering
\vspace{-1em}
\captionsetup{justification=centering}
\includegraphics[trim={9.5cm 15.5cm 10.5cm 3cm},scale=0.7]{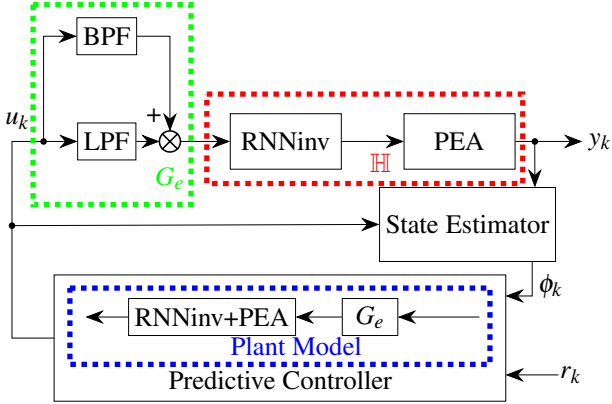}
\caption{Schematic block diagram of the RNNinv control with the proposed modeling method.}
\label{control_new}
\vspace{1em}
\end{figure}

\begin{remark}
In the case that $G$ represents the dynamics of PEA+RNNinv, it is clear that $G$ is linear if the nonlinear dynamics of the PEA is compensated by the RNNinv. Then $G_2$ is expressed in Eq. (\ref{LME}), and $G_1(j\omega)=1$ i.e., $y_{g1,k}=u_{g1,k}$. Also, Eq. (\ref{GeG}) is linear and can be further simplified. To apply the new modeling method to the controller, the control loop shown in Fig. \ref{figRNN} can be modified to that represented in Fig. \ref{control_new}. Note that LPF1 in Fig. \ref{figRNN} has been combined with $G_e$ because the cutoff frequency of LPF1 is usually much larger than that of $f_{l}$ and $f_{h}$, Thus we only need to replace HPF with a band-pass filter (BPF) with the passing frequency range of $[f_h, f_{c1}]$ ($f_{c1}$ is the cutoff frequency of LPF1 in Fig. \ref{figRNN}) as shown in Fig. \ref{control_new}.
\end{remark}

\begin{remark}
Note that the dynamics of the added $G_e$ have been included in the model for controller design, therefore, the phase lag and other dynamics of $G_e$ will not affect the tracking performance since they are accounted for by the predictive controller.
\end{remark}

\subsection{Increase the Sampling Frequency of the RNNinv Control}
If the sampling frequency of the RNNinv training set is $f_s$, then the time interval between $u_k$ and $u_{k+1}$ is $1/f_s$. Consider that the sampling frequency in real-time application becomes $2f_s$, the following state space equation can be used based on Eq. (\ref{EqRNN}).

\begin{equation}
\begin{aligned}
x_{k+2}&=\tanh(W_1x_k+B_2+B_1u_{(r),k})\\
y_{(r),k}&=W_2x_k+B_3
\end{aligned}
\label{EqRNN2}
\end{equation}
Basically, Eq. (\ref{EqRNN2}) is equivalent to two Eq. (\ref{EqRNN})s running separately  with the initial states of $x_0$ and $x_1$, respectively. $x_0$ and $x_1$ can be set to zeros at the beginning. Two sets of states, i.e., $\{x_0,x_2,x_4,\cdots\}$ and $\{x_1,x_3,x_5,\cdots\}$, are obtained separately without affecting each other. Note that for each time series, the time interval between the corresponding inputs $u_k$ and $u_{k+2}$ becomes $1/f_s$.

Next, we argue that the modeling accuracy is not affected with the sampling frequency doubled. Suppose the expected output series should be $\bar{y}_k$, if both output series $\{y_0,y_2,y_4,\cdots\}$ and $\{y_1,y_3,y_5,\cdots\}$ have the accuracy of $\epsilon$ ($\epsilon>0$), i.e., $\underset{k}{\max}~|y_{2k+1}-\bar{y}_{2k+1}|<\epsilon$ and $\underset{k}{\max}~|y_{2k}-\bar{y}_{2k}|<\epsilon$, it follows that $\underset{k}{\max}~|y_{k}-\bar{y}_{k}|<\epsilon$, which implies that the accuracy of the model Eq. (\ref{EqRNN2}) remains unchanged after the sampling frequency is increased.

\section{Stability of the Predictive Controller}
 In this work, the case of unconstrained MPC is considered in the situation where the sampling frequency is very high, like trajectory tracking of PEA systems, so adding input constraints will significantly increase the computation burden. Although faster platform based on FPGA can accelerate the computation process, it is still challenging to realize online MPC with high sampling frequency \cite{abughalieh2019survey}. Moreover, the prediction horizon should be large to ensure the closed-loop stability. This brings extra challenges in incorporating constraints in MPC even with high-speed platforms.

For the unconstrained MPC based on the linear system, by choosing sufficiently large prediction horizon, the closed-loop stability can always be achieved. The question is how to choose the smallest prediction horizon while maintaining the closed-loop stability.

We consider the MPC based on the following linear system,
\begin{equation}
\begin{aligned}
x_{k+1}&=Ax_k+Bu_k+\delta_k\\
y_k&=Cx_k   %y(k)&=Cx(k)+Du(k)
\end{aligned}
\label{lin_sys}
\end{equation}
where $\delta_k \in \mathbb{R}^{n\times1}$ represents the model uncertainty.
Let $N_p$ and $N_c$ be the prediction horizon and control horizon, respectively. The predicted control input after $N_c$ sample steps are equal, i.e., $u_{k+N}=u_{k+N+1}$ for $N>N_c$. Accordingly, we modify the reference such that $r_{k+N}=r_{k+N+1}$ for $N>N_c$ in the optimization problem.
% Note that changing the reference in this way cannot guarantee better performances, the point here is that even when the reference is changed, the analysis still works. How to modify the reference to improve the controller performance will be investigated in the future.

Let $U=[u_{k+1},u_{k+2},\cdots,u_{k+N_c}]^T$, $\Delta U=[u_{k+1}-u_k, u_{k+2}-u_{k+1},\cdots,u_{k+N_c}-u_{k+N_c-1}]^T$, $\textbf{1}_n=[1,1,\cdots,1]^T$, $\Delta_k=[\delta_k^T, \delta_{k+1}^T,\cdots, \delta_{k+N_p-1}^T]^T$ . For the linear system Eq. (\ref{lin_sys}), the predicted $N_p$ future outputs $Y$ are 

\begin{equation}
\begin{aligned}
Y&=Gx_k+H_1VU+H_2\Delta_k+Fu_k\\
&=Gx_k+H_1V(S\Delta U+\textbf{1}u_k)+H_2\Delta_k+Fu_k\\
&=Gx_k+H_1VS\Delta U+H_2\Delta_k+(F+H_1V\textbf{1})u_k
\end{aligned},
\end{equation}
with
\begin{align*}
Y&=\begin{bmatrix}
{y}_{k+1} \\
{y}_{k+2}\\
\vdots \\
{y}_{k+N_p}
\end{bmatrix}_{N_p \times 1} \hspace{-8mm}\text{,~}
G=\begin{bmatrix}
CA\\
CA^2\\
\ldots\\
CA^{N_p}
\end{bmatrix}_{N_p \times 1}\hspace{-8mm}\text{,~} 
{U}=\begin{bmatrix}
{u}_{k+1} \\
{u}_{k+2}\\
\vdots\\
{u}_{k+N_c} 
\end{bmatrix}_{ N_c \times 1} \\
H_1&=\begin{bmatrix}
0 & 0 & \dots & 0\\
CB & 0 & \dots & 0\\
\vdots & \vdots & \ddots & \vdots \\
CA^{N_p-2}B & CA^{N_p-3}B & \dots &0
\end{bmatrix} \\
H_2&=\begin{bmatrix}
C & 0 & \dots & 0\\
CA\textbf{1} & C & \dots & 0\\
\vdots & \vdots & \ddots & \vdots \\
CA^{N_p-1}\textbf{1} & CA^{N_p-2}\textbf{1} & \dots &C
\end{bmatrix}
S=\begin{bmatrix}
1 & 0 & \dots & 0\\
1 & 1 & \dots & 0\\
\vdots & \vdots & \ddots & \vdots \\
1 & 1 & \dots &1
\end{bmatrix} \\
V&=\begin{bmatrix}
I_{N_c \times N_c}\\
\begin{bmatrix}
0&\dots&0&1\\
\vdots&\ddots&\vdots&\vdots\\
0&\dots&0&1
\end{bmatrix}_{(N_p-N_c) \times N_c}
\end{bmatrix}_{N_p \times N_c}\hspace{-8mm}\text{, } 
F= \begin{bmatrix}
CB \\
CAB\\
\vdots\\
CA^{N_p-1}B 
\end{bmatrix}_{N_p \times 1}
\hspace*{-0.7cm} .
\end{align*}
The cost function for MPC is
\begin{equation}
\begin{aligned}
J&=(Y-VR_k)^T(Y-VR_k)+\rho \Delta U^T \Delta U \\
&= \Delta U^T(S^TV^TH_1^TH_1VS+\rho \textbf{I})\Delta U+2\Delta U^TS^TV^TH_1^TE+E^TE
\end{aligned}
\label{MPCJ}
\end{equation}
where $R_k=[r_{k+1},r_{k+2},\cdots, r_{k+N_c}]^T$, $\textbf{I}$ is the identity matrix, $\rho>0$, and $E=Gx_k+H_2\Delta_k+(F+H_1V\textbf{1})u_k-VR_k$. To minimize $J$, $\Delta U$ should be  

\begin{equation}
\begin{aligned}
\Delta U^*=\underset{\Delta U}{\text{argmin}}~J=-M^{-1}S^TV^TH_1^TE
\end{aligned}
\label{sol1}
\end{equation}
where $M=S^TV^TH_1^TH_1VS+\rho \textbf{I}$ is always positive definite and thus invertible. Let $E_1=[1,0,\cdots,0]$. The control law is then

\begin{equation}
\begin{aligned}
u_{k+1}=u_k+E_1\Delta U^*&=u_k-E_1M^{-1}S^TV^TH_1^TE\\
&=M_1x_k+M_2u_k+M_3R_k+M_4\Delta_k
\end{aligned}
\label{ctrl_law}
\end{equation}
where $M_1=-E_1M^{-1}S^TV^TH_1^TG$, $M_2=1-E_1M^{-1}S^TV^TH_1^T(F+H_1V\textbf{1})$, $M_3=E_1M^{-1}S^TV^TH_1^TV$ and $M_4=-E_1M^{-1}S^TV^TH_1^TH_2$.
Therefore, the original system can be augmented as

\begin{equation}
\begin{aligned}
\eta_{k+1}&=\begin{bmatrix}
x_{k+1}\\
u_{k+1}
\end{bmatrix}=\begin{bmatrix}
A &B\\
M_1 & M_2
\end{bmatrix}
\eta_{k}+\begin{bmatrix}
\textbf{0}\\
M_3
\end{bmatrix}
R_k+\begin{bmatrix}
\textbf{0}\\
M_4
\end{bmatrix}
\Delta_k\\
&=K\eta_k+\bar{M_3}R_k+\bar{M_4}\Delta_k\\
e_k&=\begin{bmatrix}
C& 0
\end{bmatrix}
\eta_k-r_k=\bar{C}\eta_k-r_k
\end{aligned}.
\label{error_dyn}
\end{equation}
Obviously, the closed-loop stability is determined by the matrix $K$. If the eigenvalues of $K$ all lie inside the unit circle, the closed-loop system is stable in the sense that if $R_k$ and $\Delta_k$ are bounded, the system output is bounded. Assuming $K$ is stable, the reference is constant, i.e., $r_k=r$, and $\Delta_k$ is constant, the steady state error can be computed as

\begin{equation}
\begin{aligned}
e_{\infty}&=\underset{N\to \infty}{\lim}[\bar{C}K^N\eta_k+\bar{C}(K^{N-1}+\cdots+I)\bar{M_3}\textbf{1}r\\
&\quad \quad \quad+\bar{C}(K^{N-1}+\cdots+I)\bar{M_4}\Delta_k-r]\\
&=\bar{C}(I-K)^{-1}\bar{M_3}\textbf{1}r+\bar{C}(I-K)^{-1}\bar{M_4}\Delta_k-r
\end{aligned}
\label{einf}
\end{equation}

If $\Delta_k$ can be neglected, a well designed control law should satisfy $\bar{C}(I-K)^{-1}\bar{M_3}\textbf{1}=1$. 

Next, we transform the error dynamics (Eq. (\ref{error_dyn})) into a SISO system with the reference signal as the sole input by assuming $\Delta_k=0$.  This enables evaluating the tracking performance of the control law (Eq. (\ref{ctrl_law})) based on the frequency response of the SISO system. Define $P_k=[r_k,~ R_k^T]^T=[r_k,r_{k+1},r_{k+2},\cdots,r_{k+N_c}]^T$. We have

\begin{equation}
\begin{aligned}
P_{k+1}=\begin{bmatrix}
r_{k+1}\\
r_{k+2}\\
\vdots\\
r_{k+N_c+1}
\end{bmatrix}&=\begin{bmatrix}
\textbf{0}_{N_c\times1} &\textbf{I}_{N_c\times N_c}\\
0 & \textbf{0}_{1\times N_c}
\end{bmatrix}
P_k+\begin{bmatrix}
\textbf{0}_{N_c\times1}\\
1
\end{bmatrix}
r_{k+N_c+1}\\
&=K_pP_k+B_pr_{k+N_c+1}
\end{aligned}.
\label{ref_dyn}
\end{equation}

With $R_k=[\textbf{0},\textbf{I}_{N_c\times N_c}]P_k=B_rP_k$, we get $\eta_{k+1}=K\eta_k+\bar{M}_3R_k=K\eta_k+\bar{M}_3B_rP_k$. Using $r_k=[1,0,\cdots, 0]P_k$, combining Eq. (\ref{error_dyn}) and Eq. (\ref{ref_dyn}) yields

\begin{equation}
\begin{aligned}
\phi_{k+1}&=\begin{bmatrix}
\eta_{k+1}\\
P_{k+1}
\end{bmatrix}=\begin{bmatrix}
K &\bar{M}_3B_r\\
\textbf{0} & K_p
\end{bmatrix}
\begin{bmatrix}
\eta_k\\
P_k
\end{bmatrix}+\begin{bmatrix}
\textbf{0}\\
M_3
\end{bmatrix}
r_{k+N_c+1}\\
&=A_{cl}\phi_k+B_{cl}r_{k+N_c+1}\\
e_k&=[C~0]\eta_k-[0~\cdots,0,1,\cdots,0]\phi_k=C_{cl}\phi_k
\end{aligned}.
\label{error_dyn_siso}
\end{equation}

For Eq. (\ref{error_dyn_siso}), the input is the reference signal $r_{k+N_c+1}$ and the output is the tracking error $e_k$. Therefore, the tracking error dynamics Eq. (\ref{error_dyn_siso}) is a SISO system with the transfer function expressed by $T(z)=C_{cl}(z\textbf{I}-A_{cl})^{-1}B_{cl}$ which describes how the reference signal is related with the tracking error.
\begin{remark}
Since it is hard to find the relationship between the eigenvalues of $K$ and $(N_p,N_c)$ at this point, we will determine the smallest prediction horizon to stabilize the system by trial and error.
\end{remark}
\begin{remark}
While the unconstrained MPC is equivalent to LQR when the prediction horizon is chosen to be infinite, the former has at least two advantages over the latter. First, LQR needs reference dynamics to perform output tracking which is not always available for arbitrary trajectories. Second, unconstrained MPC is more flexible in the sense that it can easily incorporate the uncertainty term $\delta_k$ thus quite suitable for our case (i.e., LME).
\end{remark}

\begin{remark}
Although the computation issue of the unconstrained MPC is not significant when the prediction horizon is large, we have to keep it as small as possible due to the term $\delta_k$ which is hard to be estimated in the long run. In this work, we assume $\delta_k$ is fixed for a short term considering LME is supposed to model the PEA low-frequency dynamics.
Note that there are other methods to stabilize the system, however, it may not be able to minimize the tracking error at the same time. For example, one can add another term $\rho_1U^TU$ in the cost function with large $\rho_1$ to minimize the control efforts, then the steady state error in Eq. (\ref{einf}) is not zero.
\label{whysmallNc}
\end{remark}

\section{Experimental Results and Discussion} 
In this section, first, based on the LME, the parameters of the filters and $(N_p,N_c)$ are chosen to stabilize the closed-loop system. Then, we will compare the tracking performance of the proposed method with that of an iterative learning control (ILC) approach---MIIFC \cite{kim2013modeling}.
\subsection{Choosing Modeling and Controller Parameters}
Note that the RNNinv is trained with the sampling frequency of 10kHz and is used at sampling frequency of 20kHz with the method introduced in Sec. \Romannum{3}.B.
The obtained LME model is
\begin{equation}
\begin{aligned}
\zeta_{k+1}&=\begin{bmatrix}
0.3576 &-0.3867\\
0.0537 & 1.0298
\end{bmatrix}
\zeta_{k}+\begin{bmatrix}
1.6718\\
-0.1404
\end{bmatrix}
\hat{u}_{k}+\begin{bmatrix}
1.0058\\
1.1462
\end{bmatrix}
\hat{e}_k\\
\hat{y}_k&=\begin{bmatrix}
0.4257& 0.6411
\end{bmatrix}
\zeta_k
\end{aligned}.
\label{LMEAct}
\end{equation}

Based on the modeling results of RNNinv, LME is designed to handle the system dynamics lower than 26Hz. Therefore, according to the discussions in Sec. \Romannum{2}, we set $f_l=32$Hz and $f_h=26$Hz. The second cutoff frequency of the BPF is set to be 800Hz. In terms of the orders of the filters, the higher the better. We start with 3rd order filters. All the filters are of the type Butterworth. As discussed in Sec. \Romannum{4}, the closed-loop stability is determined by the matrix $K$ and the tracking performance can be evaluated from the frequency response of Eq. (\ref{error_dyn_siso}). Note that when $N_p$ is very large, it's hard to compute the frequency response with the transfer function $T(z)$, in which case we resort to numerical approaches. For a good controller, Eq. (\ref{error_dyn_siso}) should behave like a low-pass filter with the cutoff frequency as large as possible.

\begin{figure}[t!]
\centering
\captionsetup{justification=centering}
\includegraphics[trim={0.5cm 0cm 0cm 0cm},scale=0.48]{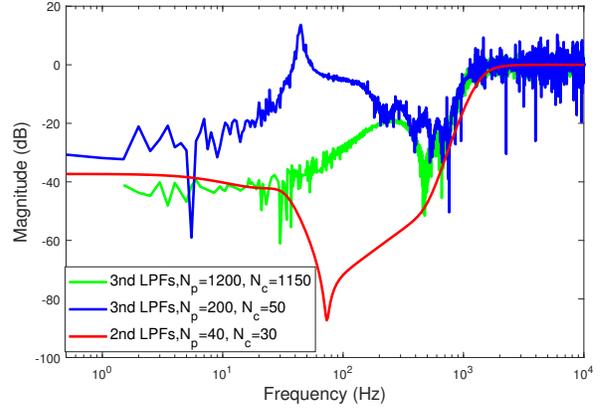}
\caption{Frequency responses of the tracking error dynamics (Eq. (\ref{error_dyn_siso})) with different ($N_p$,~$N_c$)s.}
\label{bodecmp}
\vspace{0.2cm}
\end{figure}

It turns out that if $N_p$ is close to $N_c$, the minimum $N_p$ which can stabilize $K$ is around 1200 and the frequency response of the error dynamics is shown in Fig. \ref{bodecmp}, i.e., the controller with $(N_p,N_c)=(1200,1150)$. It shows that the controlling bandwidth is about 0-1kHz. However, when $N_p-N_c$ (e.g., $(N_p,N_c)=(200,50)$) was increased, $K$ could be stabilized with a smaller $N_p$. But the corresponding control bandwidth was quite limited as seen from Fig. \ref{bodecmp}. Recall that $N_c$ should be kept small as explained in Remark. \ref{whysmallNc}. Therefore, we had to lower the orders of the filters.

\begin{figure}[t!]
\centering
\captionsetup{justification=centering}
\includegraphics[trim={0cm 0cm 0cm 0cm},scale=0.6]{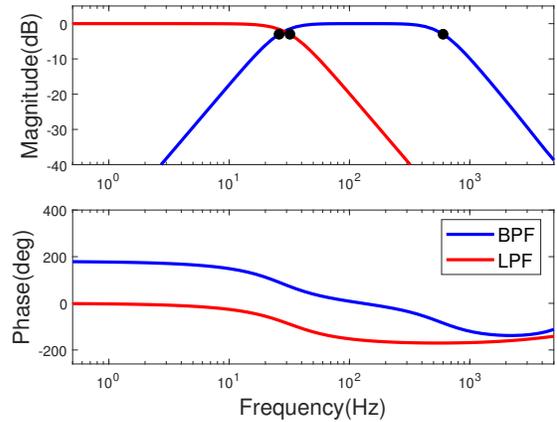}
\caption{Bode plots of the LPF and BPF.}
\label{LPFBPF}
\vspace{1em}
\end{figure}

If the filters were second order, it was easy to make the closed-loop system stable with small $N_p$ and $N_c$, and the control bandwidth could reach as high as 1kHz as shown in Fig. \ref{bodecmp}. Therefore, we chose second order filters. The price we paid for this compromise is the downgrading of tracking performance as shown later. The bode plot of the LPF and BPF are shown in Fig. \ref{LPFBPF}. Since the LPF and BPF were connected in parallel, it could be observed from Fig. \ref{LPFBPF} that all the signals can pass through the system (i.e., $G_eG$) except the signal with ultra-high frequencies (e.g., higher than 1kHz).

\subsection{Tracking Performance Comparison}
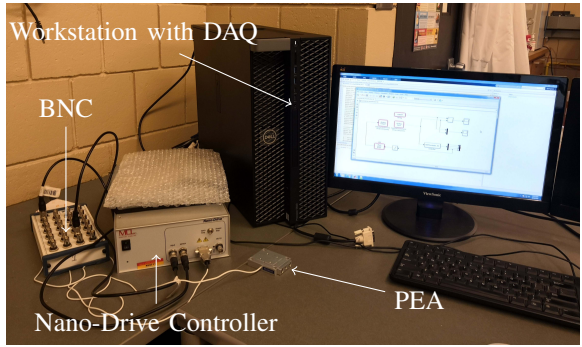
\begin{figure}[t!]
\vspace{2em}
\begin{center}
\begin{tikzpicture}
    \node[anchor=south west,inner sep=0] (image) at (0,0) {\includegraphics[scale=0.048]{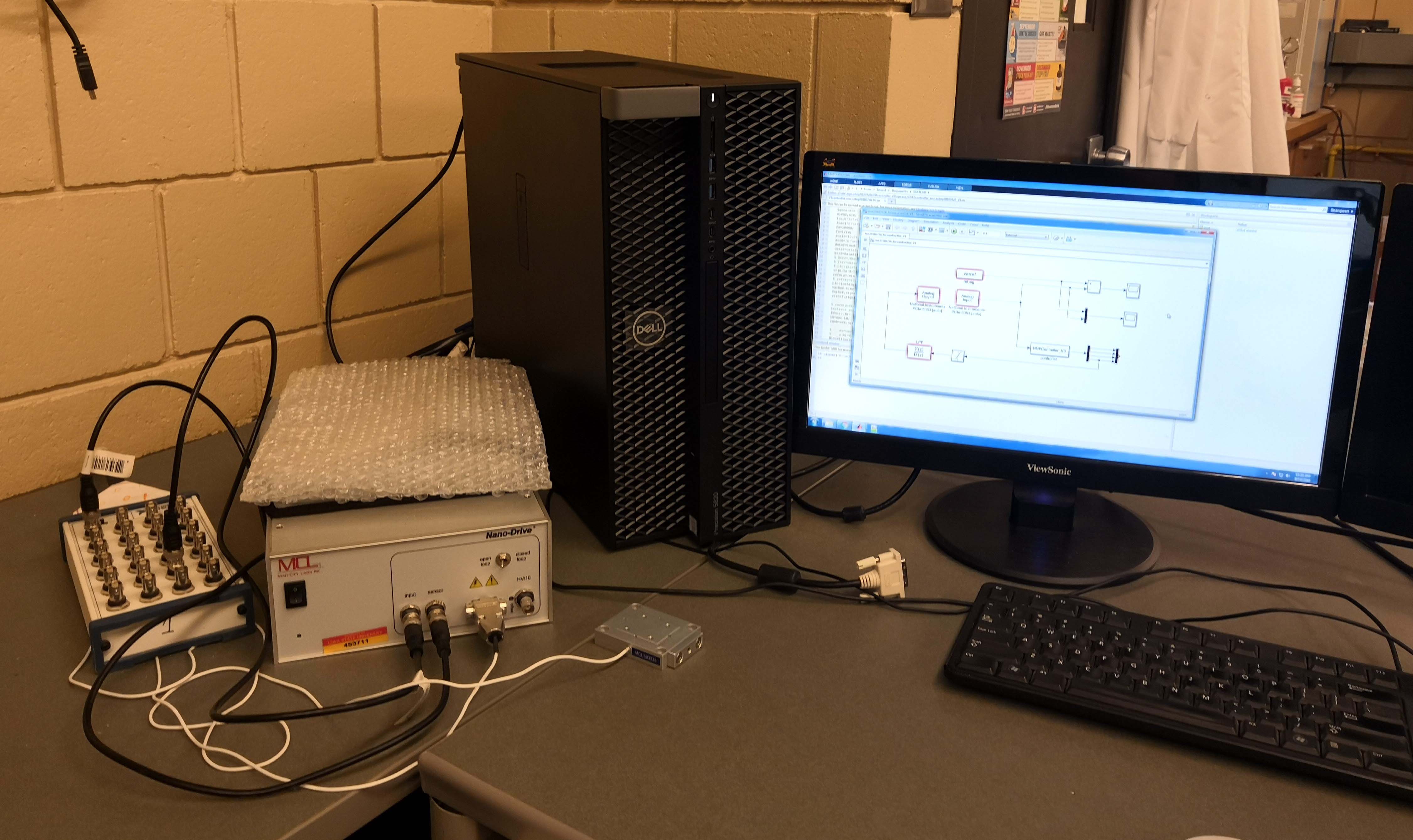} };
    \node[white] (piezo) at (5.5,0.6) {PEA};
    \node[white] (bnc) at (0.8,3.15) {BNC};
     \node[white] (controlbox) at (2,0.3) {Nano-Drive Controller};
     \node[white] (pc) at (1.7,4.15) {Workstation with DAQ};
     \draw[->,white] (piezo) -- (3.8,1.0);
     \draw[->,white] (bnc) -- (0.8,1.5);
     \draw[->,white] (controlbox) -- (2,1.2);
      \draw[->,white] (pc) -- (3.8,3.2);
\end{tikzpicture}
\end{center}
\caption{Experimental setup}
\label{expsetup}
\vspace{2em}
 \end{figure}

The proposed approach was implemented on a PEA (Nano-OP30, Mad City Labs) to track different trajectories. The tracking performance was compared with that of an efficient ILC approach---MIIFC \cite{kim2013modeling}. The experiment setup is shown in Fig. \ref{expsetup}. All the signals were collected and generated through a data acquisition system (NI PCIe-6353, National Instruments) which was installed in the workstation (Intel Xeon W-2125, RAM 32GB). The controller was designed using MATLAB Simulink (MathWorks, Inc.).

Four trajectories, including sinusoidal trajectories with the frequencies of 30Hz, 103Hz, 200Hz, and $\Gamma$ (Eq. (\ref{siggamma})) are used as the reference to be tracked.

\begin{equation}
\begin{aligned}
\Gamma(t)=[&0.35\sin(2\pi 4.2t-4.2\pi)+0.6\sin(2\pi 10.2t-10.2\pi)-\\&0.5\sin(2\pi 62.1 -2.1\pi)+0.21\sin(2\pi 127t-1.2\pi)+\\
&0.33\sin(2\pi 183t-1.3\pi)]/1.4+1.4.
\end{aligned}
\label{siggamma}
\end{equation}

The tracking errors $E_{rms}$ and $E_{max}$ were computed as follows,

\begin{equation}
\begin{aligned}
E_{max}\overset{\Delta}{=} \frac{||r(\cdot)-y(\cdot)||_\infty}{||r(\cdot)||_\infty} \text{,~~}
E_{rms}\overset{\Delta}{=} \frac{||r(\cdot)-y(\cdot)||_2}{||r(\cdot)||_2}
\end{aligned}
\label{Eqerror}
\end{equation}

where $r(\cdot)$ and $y(\cdot)$ are complex vectors obtained through discrete Fourier transform of the corresponding signals \cite{kim2013modeling}.

\begin{figure}[thpb!]
      \centering
      \vspace{0.2cm}
      \includegraphics[trim={0.8cm 0.cm 0.8cm 0cm},scale=0.6]{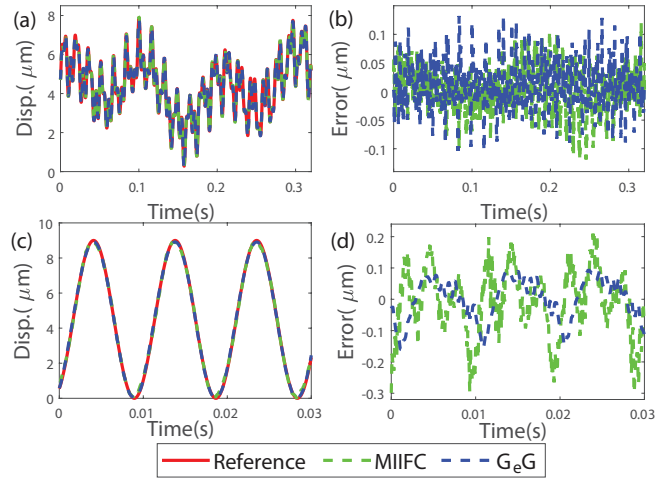}
      \caption{(a) Comparison of the tracking results for $\Gamma$ signal using $G_eG$ and MIIFC, (b) the tracking error, (c) comparison of the tracking results for 103Hz sinusoidal signal using $G_eG$ and MIIFC, (d) the tracking error.}
    \label{figCmp}
\vspace{0.3cm}
\end{figure}

Next, we use ``$G_eG$'' and ``$G$'' to denote the approaches using the method discussed in Sec. \Romannum{2} and without using the new modeling method, respectively. Table \ref{tab1} shows the tracking errors for all the three approaches while Fig. \ref{figCmp} compares the tracking performance in time domain between MIIFC and $G_eG$.

Observe that in Table. \ref{tab1}, $G_eG$ can decrease both the $E_{rms}$ and $E_{max}$ by at least 30\% for all the trajectories compared to $G$. This shows that the proposed method further improved the modeling accuracy. Compared to MIIFC, $G_eG$ outperforms MIIFC in the low-frequency region with the tracking errors about 10\% lower than that of MIIFC (for 11Hz and 103Hz cases) as shown in Table. \ref{tab1} and Fig. \ref{figCmp}. Although the tracking errors of $G_eG$ increased a little bit when tracking the 201Hz trajectory, the tracking errors are less than 5\%. This is mainly because the orders of the filters are not high enough. 

\begin{figure}[thpb!]
      \centering
      \includegraphics[trim={0.8cm 0.0cm 0.cm -0.3cm},scale=0.58]{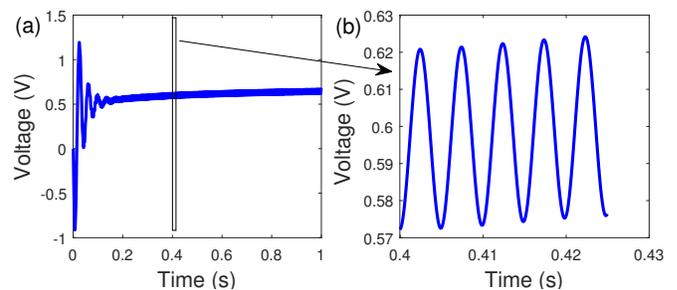}
      \caption{(a) Output of the LPF, (b) zoomed-in view of the rectangle area.}
      \label{LPFoutput}
       \vspace{0.3cm}
 \end{figure}

Here, we briefly explain how the order of the filters affect the tracking performance based on the collected data. When tracking 201Hz sinusoidal signal, the output of the LPF in Fig. \ref{control_new} should be a constant but in the experiment it includes high frequency components as shown in Fig. \ref{LPFoutput}, which then resulted in the tracking error. Based on the bode plot in Fig. \ref{LPFBPF}, about 3\% (i.e., -30dB) of the input signal can pass through the LPF which was in accordance with the experimental data considering the input amplitude was 2V. Therefore, a better LPF need to be designed with the closed-loop stability guaranteed to further improve the tracking performance.

%Without using predictive control, the tracking errors of using RNNinv alone are $E_{rms}=2.60\%$ and $E_{max}=1.79\%$ which are smaller than that of $G_eG$.

\renewcommand{\arraystretch}{1.4}
\newcommand{\len}{1.75em}
\begin{table}[thpb!]
\vspace{1em}
\caption{Tracking performance comparison of ``$G_eG$'', ``$G$'', and MIIFC when tracking different trajectories.}
  \centering
%    \begin{tabular}{c|cc|cc|cc|cc}
    \begin{tabular}{p{3.5em}|p{\len}p{\len}|p{\len}p{\len}|p{\len}p{\len}|p{\len}p{\len}}
    \hline
    Refs. &\multicolumn{2}{c|}{$\Gamma$}&\multicolumn{2}{c|}{11Hz}&\multicolumn{2}{c|}{103Hz}&\multicolumn{2}{c}{201Hz}\\
    \cline{1-9}
    $Error(\%)$&$E_{rms}$&$E_{max}$&$E_{rms}$&$E_{max}$&$E_{rms}$&$E_{max}$&$E_{rms}$&$E_{max}$\\   
    \cline{1-9}   
    $G_eG$&0.83&0.28&0.15&0.06&1.22&0.68&4.20&3.27\\
    \cline{1-9}   
    $G$& 1.14&0.39&0.24&0.09&1.93&1.10&6.35&5.33\\
    \cline{1-9} 
    MIIFC& 0.85&0.27&1.54&1.15&1.72&0.85&1.50&1.04\\
    \cline{1-9}    
    \end{tabular}
    \label{tab1}
\end{table}

 %trim={<left> <lower> <right> <upper>}

\section{Conclusion}
In this work, we enhanced the performance of RNNinv in tracking control in two aspects---the tracking accuracy is further improved and the sampling frequency is doubled. We analyzed the closed-loop stability of the control law and the equation of error dynamics is derived, which are used to guide the selection of the filter parameters and controller parameters. The proposed approach has been validated with experiments on a commercial PEA showing that the tracking precision has been significantly improved.

\section*{Acknowledgment}

This work was supported by the National Science Foundation (NSF) (CMMI-1751503) and Iowa State University.

\bibliography{ACC2020}
\bibliographystyle{ieeetr} 

\end{document}